\documentclass[twocolumn,showpacs,amsmath,amssymb]{revtex4-1}
\usepackage{hyperref}
\usepackage{graphicx}%[dvips]% Include figure files
\usepackage{dcolumn}% Align table columns on decimal point
\usepackage{bm}% bold math
\usepackage{color}

\begin{document}
\title{Scaling laws for slippage on superhydrophobic fractal surfaces}
\author{C.Cottin-Bizonne}
\email{cecile.cottin-bizonne@univ-lyon1.fr}
\homepage{http://www-lpmcn.univ-lyon1.fr/interfaces_fluides/}
\author{C. Barentin}
\email{catherine.barentin@univ-lyon1.fr}
\author{L. Bocquet}
\email{lyderic.bocquet@univ-lyon1.fr}
\homepage{http://www-lpmcn.univ-lyon1.fr/~lbocquet}

\affiliation{LPMCN;  Universit\'e de Lyon; Universit\'e Lyon 1 and
CNRS, UMR 5586; F-69622 Villeurbanne, France\\}
\date{\today}
\begin{abstract}

We study the slippage on hierarchical fractal superhydrophobic surfaces, and find an unexpected rich behavior for hydrodynamic friction on these surfaces.  We develop a scaling law approach for the effective slip length, 
which is validated by numerical resolution of the hydrodynamic equations. Our results demonstrate that slippage does strongly depend on the fractal dimension, and is found to be always smaller on fractal surfaces as compared to surfaces with regular patterns.
This shows that in contrast to naive expectations, the value of effective contact angle is not sufficient to infer the amount of slippage on a fractal surface: depending on the underlying geometry of the roughness, strongly superhydrophobic surfaces may in some cases be fully inefficient in terms of drag reduction. Finally, our scaling analysis can be directly extended to the study of heat transfer at fractal surfaces,
in order to estimate the Kapitsa surface resistance on patterned surfaces, as well as to the question of trapping of diffusing particles by patchy hierarchical surfaces, in the context of chemoreception.
%as well as 
%
\end{abstract}
\maketitle
%
%
%%%%%%%%%%%%%%%%%%%
% INTRO
%%%%%%%%%%%%%%%%%%%
%
%
\section{Introduction}
Superhydrophobic (SH)  surfaces have raised a considerable interest over the recent years \cite{Quere2005,Bocquet2011}. 
A SH surface is typically characterized by a contact angle exceeding 150$^\circ$, which results from the combination
of bare hydrophobicity and micro- and nano- scale roughness on the surfaces. This exceptional wetting property has motivated numerous studies to rationalize superhydrophobicity and device versatile SH materials, as well as to explore its implication in the context of transport and fluid dynamics 
%are interesting to change wetting properties, to reduce friction at wall and to impact transport phenomena  
\cite{Bocquet2011,Bico2002,Lafuma2003,Cottin2003,Quere2003,Ou2004,Ou2005,Joseph2006,Kim2008,Vino2010,Duez2010}. In particular, patterned SH surface were shown to exhibit low friction -- superlubricating -- properties in the Cassie state  \cite{Cassie1944}, {\it i.e.} when the liquid interface lies at the top of the roughness and thus strongly reduces the direct solid-liquid contact area \cite{PHILIP1972a, Lauga2003, Lauga2005}. This drag reduction phenomena is associated with a large slippage of the fluid at the SH surface
\cite{Ou2004,Ou2005,Joseph2006}. The quantity of interest is accordingly the effective slip length $b_\mathrm{eff}$, entering the Navier boundary condition \cite{Navier1823,Bocquet2007} for the averaged velocity field at the surface~:
$b_\mathrm{eff}\partial_z v = v_w$, where $v_w$ is the slip velocity at the wall, averaged over the lateral surface \cite{Lauga2005}.

In this paper, we study the friction properties on multiscale fractal SH surfaces. Fractal SH surfaces were indeed shown to exalt superhydrophobicity  \cite{Shibuichi1996, Onda1996}. Furthermore the fractal nature of the interface was shown to enhance the robustness of the Cassie state with respect to the Wenzel state (with the liquid impregnating the roughness), as compared to simple regular structures such as periodic stripes or pillars \cite{Jeong2008, Lee2009, Kwon2009, Liu2010}. However how the fractal, multiscale nature of the interface impacts transport and hydrodynamic slippage remains largely unknown. {Recently Feuillebois {\it et al.} could obtain rigorous bound on the slip length on Hashin-Shtrikman fractals in a thin channel geometry \cite{Feuill2009}}. Another contribution showed that for a SH interface made of pillars, randomness in the post structure does weaken the drag reduction as compared to periodic structure \cite{Samaha2011}. Also, recent experiments by Lee and Kim showed that a hierarchical surface with nanoposts lying on the top of microposts does not necessarily lead to an increase in slip length \cite{Lee2011a}. This points to the subtle connection between surface slippage and the structure of the underlying surface.  
%
%More recently drag reduction on SH surfaces made of random roughness has been investigated \cite{Samaha2011}  showing that, for a maximum allowable pressure maintaining the Cassie state, surfaces with random post distribution produce less drag reduction than those made up of periodic posts. 
%
%This first evidence of the role of the spatial distribution of a roughness pattern on drag reduction raises the question of the possible influence that more complex geometries may have. In the present work, we investigate how SH surfaces with multiscale  or fractal roughnesses  impact the friction properties. ??? Introduire angle de contact????

A key difficulty is that there is not general analytical theory describing hydrodynamic flows on composite surfaces. Therefore, analytical results for the slip length could only be obtained in specific and simple geometries with regular stripes \cite{PHILIP1972a,  Lauga2003, Lauga2005}, while asymptotic calculations on regular 2D post structures allowed to rationalize the limiting behavior of the slip length in the low surface fraction regime \cite{Lauga2010}. In this context, solving the Stokes equation with a fractal structure of boundary conditions appears hopeless. To bypass this difficulty, we resort to a scaling law approach for the fluid dynamics, which we put forward in a previous paper for regular surfaces \cite{Ybert2007}.
In particular, this allows us to derive simple scaling laws to characterize the slippage properties on hierarchical -- fractal -- surfaces  as a function of the roughness characteristics: surface solid fraction $\phi_s$ and fractal dimension $d_f$ of the interface. The validity of this approach is then validated on the basis of numerical resolution of the Stokes equations.  
%This is an effective slip length characterizing the average velocity field on the surface. 

%Using scaling iterative laws and semi-analytical calculations we calculate the effective slip length on fractal SH surfaces as a function of the solid fraction for different fractal dimensions $d_f$.
The main results of the present study are that: for 2D structures
\begin{itemize}
\item The fractal dimension $d_f$=1 corresponds to a change of behavior of the friction: for fractal dimension lower than $1$, $d_f<$1, the effective slip length diverges in the low solid fraction limit $\phi_s\rightarrow 0$, while for $d_f > $1, the effective slip length saturates to a finite value at low solid fraction. For $d_f=1$ the divergence is logarithmic as $\phi_s \rightarrow 0$.
\item For $d_f<1$ the scaling approach predicts that the effective slip length behaves as $b_{\rm eff} \sim \phi_s^{-\alpha}$ with an exponent $\alpha =({1-d_f})/({2-d_f})$ which depends directly on the fractal dimension $d_f$ of the interface. The explicit prefactor for this scaling behavior can be calculated analytically, leading to the prediction $b_{\rm eff} \simeq {3\over 8} L \left({4\over \pi}\phi_s\right)^{-\alpha}$ for circular posts and valid in the low $\phi_s$ regime. This generalizes for fractal surfaces the asymptotic result obtained by Davis and Lauga for circular posts on a square array (with $\alpha=1/2$) \cite{Lauga2010}.
\item For a given solid fraction $\phi_s$, regular surfaces (periodic stripes or pillars) lead to  higher slip lengths than for fractal surfaces. The difference between the two types of surfaces is enhanced at small solid fraction. 
\item The value of the effective contact angle $\theta_{\rm eff}$ (characterizing the wetting properties) is not sufficient to quantify the amount of slippage on the surface: the effective contact angle is not a relevant parameter to infer the slipperiness of a given surface. 
\end{itemize}

For 1D structures, the slip length for hierarchical fractal surfaces does saturate to a finite value at small surface fraction $\phi_s$ and is therefore always far below the result for regular stripes.
%As for regular periodic SH surfaces there is clearly a compromise to find between the stability and dynamical properties. \\

\par
\begin{figure}
    \centering
        \includegraphics[width=0.45\textwidth]{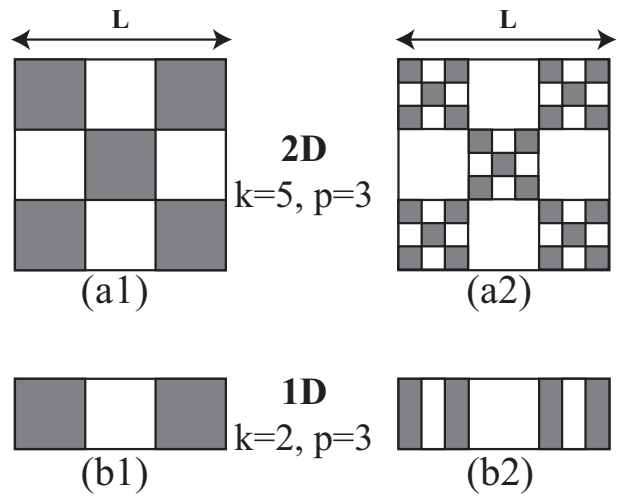}
    \caption{Iterative construction of the hierarchical surfaces considered in this study, in 2D (a1, a2) and 1D (b1, b2).
    %op view of typical 1D and 2D SH fractal surfaces considered in this paper.
    Filled patches correspond to the solid-liquid interface, while open patches correspond to the gas-liquid interface. The local boundary condition are no-slip on the filled patches and perfect slip on the open patches.
    \label{Fig1}
    In 2D for $k$=5, $p$=3, (a1) first iteration $n$=1 and (a2) second iteration $n$=2. In 1D for $k$=2, $p$=3: (b1) $n$=1 and (b2) $n$=2. }
\end{figure}

Here we consider hierarchical fractal SH surfaces obtained by the periodic repetition of a unit cell. We denote by $L$ the period of the structure. We make an iterative geometric construction of the unit cell. For the first iteration ($n$=1) the unit cell is divided in $p$ parts in each direction among which $k$ patches represent a solid surface, the other parts corresponding to the gas interface with negligeable friction. At all further iterations $n>1$, each solid patch is replaced by the pattern defined at the first iteration, with the initial period $L$ replaced by the smaller scale of the solid patch. On Fig.\ref{Fig1} we have represented two examples of the unit cells of hierarchical SH surfaces that we have studied in 2D (a) and in 1D (b) at the first iteration $n$=1 (a1 and b1) and at the second iteration $n$=2 (a2 and b2). Using this recursive construction for the hierarchical {multi-scale} structure,  there is self similarity only over a finite range of scales, the ``true'' fully fractal surface being obtained in the limit $n\rightarrow \infty$.
The fractal dimension for such structures is: 
\begin{equation}
d_f=\frac{\log k}{\log p}.
\end{equation}
  The initial solid fraction (for $n$=1) $\phi_0$, is equal  in 1D to $\phi_0=\frac{k}{p}$, and in 2D to $\phi_0=\frac{k}{p^2}$ (here patches with a squared geometry are considered). The solid fraction at the second iteration $n$=2 is then: $\phi_{s2}=(\phi_0)^2$ and more generally the solid fraction at an iteration  $n$ is: $\phi_s=(\phi_0)^n$.

\section{A Scaling law approach \label{sec:scaling}}
We first present the basis of the scaling law approach which we developed to evaluate the slip length on SH fractal surfaces. 
The starting point is a hierarchical relationship which we proposed in Ref. \cite{Ybert2007}, and showing that for a SH surface of roughness periodicity $L$, a (small) solid fraction $\phi_s$ and slip length $b_s$ on the solid at the top of the roughness, the effective slip length $b_\mathrm{eff}$ takes the following expression:
\begin{equation}\label{eq:rapPOF}
b_\mathrm{eff}\underset{\phi_s\rightarrow 0}{\simeq}b_0+\frac{b_s}{\phi_s}.
\end{equation}
In this equation, $b_0$ stands for the slip length for the same geometry when there is no slip on the solid at the top of the roughness ($b_s=0$) \cite{Note1}.
%\footnote{Note that in Ref.\cite{Ybert2007}, the prefactor in front of the second term of the above equation, with $b_s$, was incorrect by a prefactor $2\pi$ -- {\it cf.} Eq(8) and Fig. 3 in this paper --, so that the correct prefactor in front of $b_s$ is indeed close to unity.}. 
Typically we showed that $b_0\sim a/\phi_s$ where $a$ the typical lateral size of the roughness feature.

We briefly recall the physical idea underlying this relationship. It 
 is obtained by identifying the averaged friction force on the surface, which is by definition $F_s= {\cal A} \eta U/b_{\rm eff}$ -- with ${\cal A}$ the lateral area and $U$ the slip velocity --, with the total viscous force acting on the top of the solid roughness (say, posts or stripes), which writes $F_\eta = \eta \phi_s \langle \dot\gamma\rangle$, with $\langle \dot\gamma\rangle$ the average shear-rate on a single roughness feature. As proposed in Ref. \cite{Ybert2007}, a simple estimate for this quantity is $\langle \dot\gamma\rangle \sim U/(a + b_s)$, with $a$ the typical size of the solid roughness feature. This force balance leads to $b_\mathrm{eff} \sim a/\phi_s + b_s/\phi_s$, and thus to Eq.(\ref{eq:rapPOF}). 
Note that this derivation neglects 
%{Eq.}\ref{eq:rapPOF} is obtained neglecting 
hydrodynamic interactions between different roughness features and is expected to be valid in the limit of small solid fraction $\phi_s$. In Ref. \cite{Ybert2007}, we have checked in detail the validity of the above analysis and of Eq.(\ref{eq:rapPOF}). 

{Here} we extend this scaling analysis to the case of hierarchical fractal SH surfaces: the main idea is to use Eq.(\ref{eq:rapPOF}) as a hierarchical relation between the different scales of the roughness, with $b_s$ characterizing the slippage of the patch at lower scales.

\subsection{Wettability and contact angle}

Let us first consider the wettability of these surfaces. This is usually characterized by the effective contact angle $\theta_{\rm eff}$ of a droplet on the surface.  Using Cassie-law,  $\theta_{\rm eff}$ is obtained in terms of ratio between the solid-liquid and gas-liquid contact area. Introducing $\phi_s$ the solid surface fraction, one has
\begin{equation}
\textrm{cos}\theta_{\rm eff}=\phi_s(1+\textrm{cos}\theta_0)-1
\end{equation}
with $\theta_0$ the contact angle on the bare, smooth surface with the same chemical caracterisitics. As pointed above, the solid fraction is given by $\phi_s=\phi_{s1}= \phi_0$ at step $n=1$, while at step $n=2$, $\phi_s=\phi_{s2}= \phi_0^2$. For a surface characterized by a hierarchy of $n$ successive scales, then $\phi_s= \phi_{sn}=\phi_0^n$. Going up in the hierarchy, the solid fraction $\phi_s$ thus decreases  and the contact angle on those hierarchical fractal surfaces  converges {\it exponentially} towards $180^{\circ}$: 
%\subsubsection{Contact angle}
%A current way to characterize a SH surface is to look at its wettability characterized by  $\theta_{eff}$, the effective contact angle of a drop of a liquid placed on the surface. This angle depends on the chemical nature and on the roughness of the surface. 
%The effective contact angle $\theta_{eff}$ on a hierachical fractal surface at a step $n$ can be explained as a function of $\theta_0$ -the contact angle on a smooth surface with the same chemical caracterisitics- and the solid fraction $\phi_S=(\phi_0)^n$ :
%\begin{equation}
%\textrm{cos}\theta_{eff}=(\phi_0)^n(1+\textrm{cos}\theta_0)-1
%\end{equation}
%Such that,
\begin{equation}\label{thetaeff}
\textrm{cos}\theta_{\rm eff}+1 \underset{n \rightarrow \infty}{\rightarrow} 0
\end{equation}
%The contact angle on those hierarchical fractal surfaces quickly  converges toward $180^{\circ}$.

Let us now turn to the slippage properties.
\subsection{Slippage on a two scale roughness}
To illustrate the approach, 
we begin by studying the simplest  hierarchical surface {obtained with} two steps in the iteration process ($n$=2), i.e., two roughness scales as represented in Fig.\ref{Fig1} (a2) and (b2). Such hierarchical surfaces are now accessible {from} experiments {thanks to} the progress in nanofabrication {techniques} \cite{Jeong2008, Kwon2009}.

Let us introduce $b_1$, the effective slip length obtained 
after one iteration $n$=1, see Fig.\ref{Fig1} (a1) and (b1). A key remark is that,  when going from the first $n=1$ to the second step $n$=2, the higher-level structure inside each main solid patch is equivalent to the initial bare structure at $n=1$, up to a rescaling of the size by $1/p$ ($p$ being the size ratio between two scales, as defined above). 
%Accordingly, the higher-level structure inside each patch as in Fig.\ref{Fig1}(a1) is characterized by 
%slip length $b_s=b_1/p).   characterizing the underlying the are characterized by a slip length $b_1/p$ (the slip length scales up with the periodicity of the system)

Using Eq.\ref{eq:rapPOF},
%and the fact that at the second step $n$=2 the solid patches are characterized by a slip length $b_1/p$ (the slip length scales up with the periodicity of the system), 
we then deduce that the effective slip length is equal to:
\begin{equation}\label{eq:rapPOF2steps}
b_2=b_1+\frac{b_1}{p\phi_0},
\end{equation}
where we recall that $\phi_0$ is the solid fraction at the initial scale, in Fig.\ref{Fig1}(a1)-(b1).
For 1D structures like in Fig.\ref{Fig1}(b1), $\phi_0=k/p$, so that one obtains 
\begin{equation}\label{eq:rapPOF1D}
b_2=b_1\times(1+\frac{1}{k}),
\end{equation}
For 2D structures like in Fig.\ref{Fig1}(a1), $\phi_0=k/p^2$ and
\begin{equation}\label{eq:rapPOF2D}
b_2=b_1\times(1+\frac{p}{k}),
\end{equation}
The slip length is obviously higher at the second iteration. There is an increase in slip length for  hierarchical ``fractal'' SH surfaces and this increase is larger for 2D structures than 1D.
Such a difference between the 1D and 2D structures was already pointed out for regular periodic SH surfaces \cite{Ybert2007}.

%%%%

%
\subsection{Slippage on hierarchical fractal surfaces}
Now, ``real'' SH fractal surfaces are composed by an infinite number of iterations, or at least by a large number $n$ of it, so that the fractal dimension is defined over a broad range of scales. To estimate the slip length, we extend the previous analysis for any number of  iterations. 

%\subsubsection{Slip length}
At a subsequent iteration $n+1$, the solid patches are divided once again into $p$ parts, see Fig.\ref{Fig1}. As an alternative view, one may consider that the system at the subsequent iteration $n+1$ is made by first assembling $k$ systems obtained at the previous iteration $n$ in order to gather a structure with the correct geometry, and then perform a global rescaling by $1/p$ to obtain the desired structure at step $n+1$.  Accordingly, the structure inside the largest patches at the iteration $n+1$ can be seen as the previous structure at iteration $n$, up to a rescaling factor $p$ between scales. 

One may deduce directly the recursion relationship for the slip length from the above procedure. At scale $n+1$ the slip length over each of the largest patches is then $b_n/p$ (since we have gathered structures from iteration $n$ and rescaled by $1/p$; remind that the slip length does scale with the global lateral size of the system). 
Now
  using Eq. \ref{eq:rapPOF}, the slip length $b_{n+1}$  at iteration $n+1$  can be deduced as a function of $b_{n}$ as:
\begin{equation}
%b_{n+1}=b_{n}(1+x)\ \textrm{with} \ x=\frac{1}{p\phi_0}. 
b_{n+1}=b_1+\frac{1}{\phi_0}\times {b_n\over p}
\end{equation}
The last term thus gathers all the slippage effects on all above scales $\ge n$ in the hierarchy; $b_1$ is the slip length of the elementary structure, as in Fig.\ref{Fig1}(a1)-(b1).
%Since the solid patches at scale $n$ are just rescaled version up to a proportionality factor $p$ of the system at smaller scales, then one deduces that the slip length $b_s$ on the solid patches at scale $n$ can be written as  $b_{s}=b_n/p$ :
%(the slip length scales up with the periodicity of the system), that 
%We deducecan be written as
This can be rewritten as 
\begin{equation}\label{eq:incrfrac}
b_{n+1}=b_1+b_{n}x\ \ \textrm{with} \ \ x=\frac{1}{p\phi_0}. 
\end{equation}
so that
\begin{equation}
\label{bn}
b_{n}= b_1 \times \sum_{k=0}^{n-1} x^k= b_1\frac{1-x^n}{1-x}
\end{equation}
%with 
%\begin{equation}\label{eq:x}
%x=\frac{1}{p\phi_0}. 
%\end{equation}

%For $x\ne1$ $b_n=b_1\frac{1-x^n}{1-x} $ 

Depending on the parameter $x$, different scenarios occur:
\begin{itemize}
\item For $x<1$, the sum in Eq.(\ref{bn}) converges exponentially  to 
a finite value in the limit  $n\rightarrow \infty$:
%the effective slip length at an iteration $n$ is given by: 
%\begin{equation}\label{eq:xinf1}
%b_n=b_1\frac{1-x^n}{1-x} . 
%\end{equation}
%, the effective slip length converges to a finite value: 
\begin{equation}\label{eq:xinfin}
b_\infty=b_1\times \frac{1}{1-x}
\end{equation}
\item For $x=1$, the effective slip length at an iteration $n$ is proportional to $n$, $b_n= n\times b_1$.
Now, replacing the scale index $n$ by the corresponding solid surface $\phi_{s}$ at the corresponding
scale, $\phi_s=\phi_0^n$ so that $n=\log\phi_s/\log\phi_0$, then one gets
\begin{equation}\label{eq:xeq1}
b_{\rm eff}(\phi_s)= b_n=b_1\times \frac{\log\phi_s}{\log\phi_0}.
\end{equation}
The slip length thus diverges  \textit{logarithmically} with the solid fraction $\phi_s$ in the limit of small $\phi_s$.

\item For $x>1$, the largest term in Eq.(\ref{bn}) dominates the effective slip length, so that
the slip length has the asymptotic expression in the limit $n\rightarrow \infty$ ($\phi_s\rightarrow 0$)
%at an iteration $n$ is given by:
%\begin{equation}\label{eq:xsup1}
%b_n=b_1(\sum_{i=0}^{n-1} \ x ^{i})=b_1\frac{1-x^n}{1-x} .
%\end{equation}
% Therefore, in the limit of high values of $x$ ($x\gg1$), the slip length has the asymptotic form:
\begin{equation}\label{eq:bpuisdf}
b_{\rm eff}(\phi_s)= b_n\simeq \frac{b_1}{x-1}\times x^{n},
\end{equation}
with $n=\frac{\log\phi_s}{\log\phi_0}$.
\end{itemize}

%%%
{As a side note, one may remark that these results could be generalized to the situation with a finite slip length $b_0$ on the liquid-gas interface. Our present results thus propose an upper bound for the effective slip length in the latter situation. We leave this generalization for future work.
}

\subsection{Discussion: 1D/2D hierarchical surfaces versus regular surfaces}
We now summarize the different scaling expression for the slip length for 1D and 2D structures and for the different values of $x$. A key quantity is the fractal dimension of the structure, $d_f=\frac{\log k}{\log p}$. \\
%\begin{itemize}
\noindent$\bullet$  For 1D structure, $x=1/k$ and $k<p$, so that $x\leq1$ and $d_f<$1.\\
\noindent$\bullet$  For 2D structure, $x=p/k$ and $k<p^2$ so that one finds three different cases: for $x<$1, $1<d_f<2$; for $x=1$,  $d_f$=1; and for $x>$1, $d_f<$1.
%\end{itemize}

\subsubsection{1D structures}
In 1D, two situations occur:
\vskip0.3cm
%\begin{itemize}
%\item 
\noindent$\bullet$ For 0$<d_f<$1 ($x<$1),  then $b_\infty=b_1\frac{k}{k-1}$, and the maximum value of ${b_\infty}/{b_1}$ is 2 (for $k=2$). This shows that the hierarchical fractal geometry only generates a maximal amplification by a factor of  two for the slip length as compared to a regular SH surface defined with only one roughness scale. In this regime, the slip length does not diverge in the limit of perfect superhydrophobicity, $\theta_{\rm eff}=180^\circ$, or $\phi_s=0$.\\

%\item 
\noindent$\bullet$ The peculiar case $d_f$=0 ($x=1$, {\it i.e.} k=1) corresponds to a flow parallel to stripes.  We recover in this case the  slow logarithmic divergence of the slip length with the solid fraction $b_n \sim{\log\phi_s^{-1}}$ as obtained in Refs. \cite{PHILIP1972a}. This result is non trivial as in the present approach we do not expliclity take into account the hydrodynamics interactions between stripes.
%\end{itemize}

\subsubsection{2D structures}

In 2D, a variety of situations occurs depending on the ration $x=p/k$. The latter parameter quantifies the dense or sparse character of the solid filling on the surface.
\vskip0.3cm
%\begin{itemize}
%\item 
\noindent$\bullet$ For $1<d_f<2$ ($x=p/k<1$), the solid filling is quite dense -- this is the situation depicted in Fig.\ref{Fig1}(a1)-(a2)  --. In this regime, the slip length of the SH hierarchical surface saturates to a value   $b_\infty=b_1\frac{k}{k-p}$ in the limit
of low $\phi_s$.
%The latter is larger than the saturation value for 1D structures $b_\infty = b_1\frac{k}{k-1}$. 
%Note that this value is higher than its counterpart in 1D.

%\item 
\noindent$\bullet$ For $d_f=1$ ($x=p/k=1$), we find a logarithmic divergence of the slip length with the solid fraction $b_n=n\times b_1=\frac{\log\phi_s}{\log\phi_0}b_1$. We find the same divergence as for a 1D regular SH surface of stripes.

%\item 
\noindent$\bullet$ For $d_f<$1 ($x=p/k>$1), the solid filling is sparse. In this case the effective slip length diverges as a power law of the solid fraction. Using Eq.\ref{eq:bpuisdf}, we get for large $n$, {\it i.e.} small $\phi_s$: 
\begin{equation}\label{eq:div2D}
b_{\rm eff}(\phi_s)=b_nÊ\approx \frac{1}{x-1}b_1\times \phi_s^{-\frac{1-d_f}{2-d_f}}
\end{equation}
with again $d_f=\frac{\log k}{\log p}$ the fractal dimension.
%\end{itemize}

We can compare the scaling behavior predicted in this last regime with the one obtained for a  SH surface made out of a bidimensional regular pattern of posts for which the slip length behaves for small $\phi_s$ as $b_{\rm eff} \sim \frac{L}{\sqrt{\phi_s}}$ \cite{Ybert2007}. Since $\frac{1-d_f}{2-d_f}<\frac{1}{2}$, this shows that for a given solid fraction $\phi_s$, the slip length on the periodic patterns of post is larger than of 2D SH  fractal surfaces. This is a counter-intuitive result, which shows that fractality, though strongly increasing the non-wettability, is far less efficient in enhancing dynamic properties.

\subsection{Alternative scaling law approach in the dilute regime}

Here we propose a simpler, alternative description to account for the scaling laws above. We focus on the case of a sparse solid filling in 2D, corresponding to $d_f <1$ and $x$ larger than unity.

Then in the dilute regime, when the hydrodynamics interactions can be neglected,   the effective slip length ${b}_\textrm{{eff}}$ can be estimated directly from the calculation of the viscous friction force $F$ of a liquid of viscosity $\eta$
 on a unit cell of size $L$ of the hierarchical SH surface at a given  step $n$. We consider a flow leading to a slip velocity $U$ at the surface. By definition of the slip length, the friction force on the surface takes the expression:
\begin{equation}
F=\eta\,  \frac{{U}}{{b}_{{\rm eff}}}\, L^2
\end{equation}
with $\eta$ the shear viscosity of the fluid.
Now the friction force can be calculated as the sum of the forces on each solid post. Separating the scale of the post from the lateral scale between the post, this friction force can be calculated from the Stokes equation applied to the posts considered as  individual stationary solid object in a fluid moving at velocity $U$ far from it:
\begin{equation}
F_\nu=\sum_{solid\ patches} \chi\, \eta\, a_n\,{U} 
\label{Fnu}
\end{equation}
Here  the sum runs over all the solid patches on the surface, with characteristic lateral size $a_n$ at the smaller scale $n$; here $\chi $ is the numerical prefactor for the Stokes law. The latter depends only on the geometrical characteristics of the solid post. For example for a solid disk with radius $a$, one has $\chi = {1\over 2} \times {32\over 3}$
\cite{Happel}. The factor ${32\over 3}$ comes from the Stokes flow calculation around a disk, as calculated in Ref.\cite{Happel}, while the $1/2$ prefactor stems from the fact that only the upper-half of the disk undergoes friction.

Now at an iteration $n$, the lateral size of the solid post scales like $a_n \sim L/p^n$ and there are $k^n$ of such posts.
%and $s$ is the lateral size $s=L\times \left(\frac{1}{p}\right)^n$ of all 
%the $k^n$ square solid patches. 
Altogether equating the two expressions for the friction force, we thus obtain the scaling law for the slip length in the dilute regime:
\begin{equation}
{b}_\textrm{{eff}} \sim {L}\left(\frac{p}{k}\right)^n
\end{equation}
Using the definition of $\phi_s$ and $d_f$ this leads to the scaling behavior
\begin{equation}
{b}_\textrm{{eff}} \sim \phi_s^{-\frac{1-d_f}{2-d_f}}
\label{befffrac}
\end{equation}
and we recover the same result  as in Eq.\ref{eq:div2D}, obtained with the previous hierarchical approach.

Now we finally remark that further insight can be obtained from the above analysis to predict the prefactor of the previous scaling behavior. We will focus on posts with disk shape for which the prefactor $\chi$ in the viscous force, in Eq.(\ref{Fnu}), can be exactly calculated, as pointed out above. 

We first start by noting 
%As a side remark, we note 
that applying a similar analysis to a {\it regular} square pattern of disk-like posts yields $ b_{\rm eff}= 3/16\times L^2/a$ with $a$ the disk radius, and we used $\chi =16/3$ for the disks as discussed above. Using $\phi_s=\pi a^2/L^2$ for the surface fraction, one gets 
the following expression for the slip length
\begin{equation}
b_{\rm eff} \simeq {3\sqrt{\pi}\over 16} {L\over \sqrt{\phi_s}}
\label{beffscal}
\end{equation}
The prefactor ${3\sqrt{\pi}\over 16}$ is the {\it exact one} calculated by David and Lauga \cite{Lauga2010} using asymptotic expansions and confirmed in numerical resolution of the hydrodynamic problem \cite{Ybert2007}. This demonstrates the power of the scaling approach in deriving relevant expression for slip lengths.
The above analysis can also immediately extended to predict the slip length on an array of posts of any shape ({\it e.g.} ellipsoids) and any pattern symetry.

Coming back to the hierarchical pattern, a similar analysis for disk-like posts allows to show that 
$b_{\rm eff}/L=2/\chi \times x^n$ with $\chi =16/3$ for disks. Using $\phi_s={\pi\over 4} (k/p^2)^n$, where
the $\pi/4$ factor takes into acount the disk shape of the posts, this equation can be rewritten as
%provides the prefactor for Eq.(\ref{befffrac})
%which can be rewritten as
\begin{equation}
{b}_\textrm{{eff}} \simeq  {3\over 8} L \left({4\over \pi}\phi_s\right)^{-\frac{1-d_f}{2-d_f}}
\label{beffbis}
\end{equation}
%with $\phi_0=\pi R^2/L^2$ the initial solid fraction now specified to a disk shape. 
The previous result for a regular square pattern of
disk posts is exactly recovered by putting $d_f=0$ in this equation.

Furthermore this result can be rewritten as $b_{\rm eff}=b_1 \times x^{n-1}$, with $b_1={2L\over \chi} x$: this expression is identical to Eq. (\ref{eq:bpuisdf}) obtained by the alternative iterative derivation, for $x\gg 1$ (sparse structure).
%$= b_1/x \times ({4\over \pi} \phi_s)^{-\alpha}$ with $\alpha=({1-d_f})/({2-d_f})$. 
%Up to the factor $4/\pi$ which stems from the disk shape of the posts considered here, this result is
% identical to Eq.(\ref{eq:div2D}) obtained by the alternative iterative derivation,  for $x\gg 1$ (sparse structure). 
This demonstrates the equivalence between the direct and hierarchical calculations for the slip length on fractal surfaces.

\section{Numerical results}

In order to assess the validity of the above scaling approach, we perform a numerical resolution of the full hydrodynamic equations complemented by the composite boundary conditions on the hierarchical surface, as sketched for example in Fig.\ref{Fig1}. Specifically, the hydrodynamic velocity field obeys Stokes equation in the bulk, with a no-slip boundary condition on the solid patches and a perfect slip boundary condition on the remaining liquid-gas interface (open patches in Fig.\ref{Fig1}).

%We have checked the validity of the above iterating approach using a numerical calculation that takes into account the hydrodynamic interactions for the evaluation of slip length.
To this end, we follow the numerical approach developped by Cottin-Bizonne {\it et al.} in Ref.
\cite{Cottin2004}. We only recall here the basic steps of this
approach, and a more detailed description can be found in \cite{Cottin2004}.
A shear flow is considered over a composite surface characterized by a
heterogeneous slip length pattern. The boundary is
modeled by a pattern of local slip lengths on a planar surface. The hydrodynamic equations
for flow profile are rewritten analytically in terms of a boundary integral method, which is then solved
numerically. The effective slip length is deduced from the asymptotic flow profile far from
the surface. 
%The characteristics of the flow far away from the surface
%and an effective slip length are determined using an integral method
%by solving the hydrodynamic equations with the
%hydrodynamic boundary condition given by the local slip length.

\par
Using this numerical approach, we calculate the effective slip length $b_{\rm eff}$
for SH hierarchical fractal surfaces as the ones represented on Fig.\ref{Fig1}. The computed slip lengths $b_\mathrm{eff}/L$ will be 
calculated for different SH hierarchical surfaces, for both 1D and 2D structures, as well as for different values of $k$ and $p$ and   iteration steps, $n$. The numerical values will then be compared with the predictions obtained using the scaling law approach.

In practice, due to the large range of spatial length scales involved, the numerical calculations require a higher discretization than the one performed with a regular square lattice of solid patches as considered in 
\cite{Ybert2007}. The calculations involve subsantially larger matrices, which limits the range of fractal surfaces that can be considered. Consequently,  the numerical method is limited  to only a few iterations as shown in Fig.\ref{Fig3} and \ref{Fig5}. This will however prove sufficient to explore the main tendencies predicted by the hierarchical approach.
%due to numerical limitations inherent in the integral method.

\subsection{Numerical results for slippage over hierarchical 2D structures}

\begin{figure}
    \centering
        \includegraphics[width=0.45\textwidth]{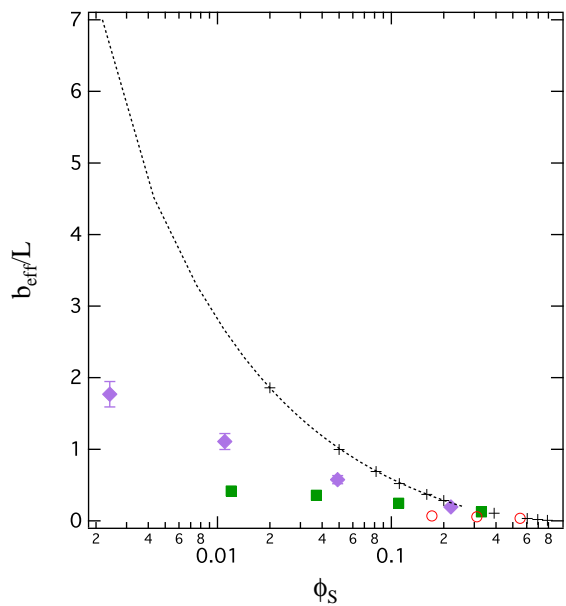}
    \caption{Numerical results for the normalized effective slip length $b_\mathrm{eff}/L$  on 2D structures, as a function of the solid fraction $\phi_s$ (in log scale) for posts ($+$) and hierarchical surfaces ($\blacklozenge$:\{$k$,$p$\}=\{2,3\}, $\blacksquare$: \{$k$,$p$\}=\{3,3\}, 
    $\circ$:\{$k$,$p$\}=\{5,3\}). The  dotted line corresponds to the scaling law for regular posts from \cite{Ybert2007}. Typical error bars are represented  for \{$k$,$p$\}=\{2,3\}.}
    \label{Fig2}
\end{figure}

We have first plotted in Fig.\ref{Fig2} the numerical results for the effective slip length $b_\mathrm{eff}/L$ as a function of the solid fraction $\phi_s$, for SH surfaces made out of both periodic posts on a square lattice and hierarchical fractal surfaces obtained with different parameters $k$ and $p$. A first striking result emerging from this plot is that the surfaces with regular structures of posts are far better than all the fractal surfaces in terms of slip length for a given solid fraction. 
%For a given $\phi_s$ the fractal surfaces are less efficient for slippage than surfaces with posts.

\begin{figure}
    \centering
        \includegraphics[width=0.45\textwidth]{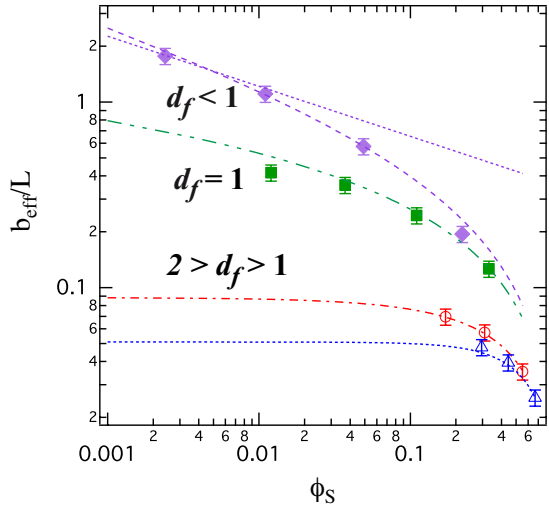}
    \caption{{Comparison between scaling law predictions and numerical results  for 2D structures}: normalized slip length $b_{\rm eff}/L$   as a function  of the solid fraction $\phi_s$  (in a log-log plot) for hierarchical fractal surfaces with different fractal dimensions $d_f$. Theoretical predictions of the scaling approach, Eq. (\ref{bn}), are shown as lines for the various fractal dimensions $d_f$ under consideration (see text for details).  Numerical results are from top to bottom: (i) $d_f<$1: \{$k$,$p$\}=\{2,3\} ($\blacklozenge$); (ii)  $d_f=$1: $\blacksquare$\{$k$,$p$\}=\{3,3\}; (iii) $d_f>$1: $\circ$\{$k$,$p$\}=\{5,3\} and $\triangle$\{$k$,$p$\}=\{6,3\}. The top dotted line is the predicted asymptotic scaling in Eq.(\ref{beffbis}), $b_{\rm eff} \simeq {3\over 8} L \left({4\over \pi}\phi_s\right)^{-\alpha}$ 
    with $\alpha=(1-d_f)/(2-d_f)$. Here $\alpha \simeq 0.27$ for the fractal dimension $d_f=0.63$ corresponding to the hierarchical surface with \{$k$,$p$\}=\{2,3\}.
    %{\color{red} figure scaling resuls a refaire}
%
%     \textit{Numerical results}: $d_f<$1: $\blacklozenge$ \{$k$,$p$\}=\{2,3\} ; $d_f=$1:$\blacksquare$\{$k$,$p$\}=\{3,3\}; $d_f>$1: $\circ$\{$k$,$p$\}=\{5,3\}, $\triangle$\{$k$,$p$\}=\{6,3\}. 
%      \textit{Scaling approach}: For $d_f>$1 the slip length saturates, (...):   \{$k$,$p$\}=\{6,3\}  and (-.-): \{$k$,$p$\}=\{5,3\}. For $d_f$=1, $\square$: \{$k$,$p$\}=\{3,3\} calculated from $\phi_{s2}=0.11$. For $d_f<$1 there is a divergence of the slip length, $\diamond$: \{$k$,$p$\}=\{2,3\} calculated from $\phi_{s2}=0.05$. The black dashed lines are just guides for the eye. \\
%      {\color{red} Pourquoi on ne peut pas estimer le reste des courbe theoriques a partir d'une valeur de reference pour la prediction ?} 
      }\label{Fig3}
\end{figure}

The detailed comparison between the numerical results and the incremental scaling laws is then shown in Fig.\ref{Fig3}.  A general conclusion from this plot is that the scaling predictions in the previous section are in  good agreement with numerical results. 
%where $\mathrm{log}(b_\mathrm{eff.}/L)$ is represented as a function of $\mathrm{log}(\phi_S)$ for fractal surfaces (and different couples $k$ and $p$). 

\par
\noindent {\it {(i)} Case $x=p/k<1$ {\it i.e.} fractal dimension $d_f>$1}: two cases have been solved numerically with hierarchical structures corresponding to \{$k$,$p$\}=\{5,3\} and \{$k$,$p$\}=\{6,3\}. At the first iteration $n=1$, these structures have a surface fraction of $\phi_0 = 0.55$ and $\phi_0 = 0.66$ respectively. For further iterations, the solid fraction obeys $\phi_s(n)=\phi_0^n$.

The scaling predictions in this case suggest a finite slip length, which saturates at small $\phi_s$. This is indeed the tendency shown by the numerical results. 

To be more quantitative, we use the results for the slip length in  Eq.\ref{bn}: $b_n=b_1({1-x^n})/({1-x}) $ with $n={\log(1/\phi_s)}/{\log(p^2/k)}$.
One has to fix the numerical prefactor  $b_1$ in the previous prediction: here we choose to calculate this value numerically for a surface structure at step $n=1$, such as on Fig.\ref{Fig1}(a1). 
%
%  using the incremental law and Eq.\ref{bn} the slip length at the iteration $n$ can be expressed as a function of $\phi_s$: $b_n=b_1\frac{1-x^n}{1-x} $ with $n=\frac{\log(1/\phi_s)}{\log(p^2/k)}$ and $x=p/k$. 
%We observe a saturation of the effective slip length for small $\phi_s$. We can also estimate the value at saturation: $\frac{b_1}{1-x} $. 
Using this calculated parameter as an input, 
a quantitative agreement is then found between the scaling prediction and the numerical results, as shown in Fig.\ref{Fig3}.
%%%
%%%
\vskip0.3cm
\noindent{\it (ii) Case $x=k/p=1$, {\it i.e.} fractal dimension $d_f=$1}: the numerical calculations have been performed for a hierarchical structure with \{$k$,$p$\}=\{3,3\}. At the first iteration $n=1$, this structure has a surface fraction of $\phi_0 = 0.33$.
Note that only four iterations could be calculated numerically.
%{\color{red} 0.22 ou 0.33, cf figure 3}. 

The scaling laws predict in this case a logarithmic dependence on the solid fraction $\phi_s$ and we thus compare the numerical expression with the estimate $b_n \sim {\log(\phi_s^{-1})}$. 

As above, we compute the numerical prefactor for this behavior by calculating the value of the slip length for a given iteration. 
%We calculated numerically $b_2$  at the $n=2$ iteration (with $\phi_{s2} = 0.11$) 
%to fix the prefactor of this $\phi_s$ dependence. The predicted slip length takes the following expression in this case
%$b_{\rm eff}(\phi_s)=b_2\times n/2$, with $n=\log(\phi_s)/\log(\phi_0)$.
We calculated numerically $b_1$  at the $n=1$ iteration (with $\phi_{s1} = 0.33$) 
to fix the prefactor of this $\phi_s$ dependence. The predicted slip length takes the following expression in this case
$b_{\rm eff}(\phi_s)=b_1\times n$, with $n=\log(\phi_s)/\log(\phi_0)$.

%{\color{red}Ên=2 vraiment necessaire ici ?}

Using this value as an input, 
we obtain a quantitative agreement between the predictions and the numerical results 
% To be quantitative, we use as a prefactor for the previous prediction the value of the slip length measured numerically for the $n=2$ iteration, $b_2$, and corresponding to a solid fraction $\phi_{s2} = 0.11$.
%$\times b_2/{\log(\phi_{s2})}$.  We considered the structure for the $n=2$ iteration as a reference for the prefactor, corresponding to a solid fraction $\phi_s = 0.11$.
%Again, there is a good agreement between the scaling law and the numerical result 
for all $\phi_s \le 0.1$, demonstrating the robustness of the scaling approach up to relatively high solid fraction.
%We believe that the small discrepancy comes from the fact that the incremental law does not take into consideration the hydrodynamics interactions between posts that plays an increasingly important role for larger values of $\phi_s$. %but are of course included in the numerical resolution. 

%Note that the prefactor for the scaling result for the slip length could be similarly evaluated from the $n=1$ iteration with $b_1$ using Eq.\ref{eq:xeq1}. In this case the agreement is slightly less good because the starting solid fraction  ($\phi_{s1}=0.22$) is too high for the scaling arguments to apply quantitatively.  
%the incremental law is represented for $\phi_s \le 0.11$ considering $b_2$ and $\phi_{s2}$ as the input parameters; $b_n$ is given by: $b_n=\frac{\log(\phi_s)}{\log(\phi_{s2})}b_2$. 
%%%
%%%
\vskip0.3cm
 \noindent{\it (iii) Case $x=p/k>$1, {\it i.e.} fractal dimension $d_f<$1}: the numerical calculations have been performed for a hierarchical structure with \{$k$,$p$\}=\{2,3\}, with $d_f=\log k/\log p=0.63$. At the first iteration $n=1$, this structure has a surface fraction of $\phi_0 = 0.22$.
Note that only three iterations could be calculated numerically for this very sparse system.

In the low $\phi_s$ regime, the hierarchical approach predicts a slow divergence {\it vs} surface fraction,  $b_{\rm eff} \sim \phi_s^{-\alpha}$ with $\alpha={{(1-d_f)}/ {(2-d_f)}}\simeq 0.27$. %and $d_f=\log k/\log p$.
%Here $d_f=0.63$ and the exponent $\alpha$ is relatively small, $\alpha\simeq 0.27$. 
%The numerical results are in semi-quantitative agreement with this scaling relationship. 
To be more precise, we show in Fig.\ref{Fig3} (top dotted line)
the analytical asymptotic prediction obtained in Eq.(\ref{beffbis}),  $b_{\rm eff} \simeq {3\over 8} L \left({4\over \pi}\phi_s\right)^{-\alpha}$ with $\alpha=(1-d_f)/(2-d_f)$.  This expression provides a result in good agreement with numerical results for small $\phi_s$ \cite{Note2}.
%{\color{red} commentaire sur disk vs square ? comment ?}
%\footnote{We note as a side remark that the numerical results are obtained for square-like posts, while the estimate of the prefactor  is obtained for disk-like posts, in particular for the prefactor of Stokes law for disks, $\chi=16/3$. However, these two estimates should differ only by a small amount: in Ref. \cite{Ybert2007}, the numerical prefactor for $b_{\rm eff}/L$ {\it vs} $\phi_s^{-1/2}$ for regular square-like posts structures was found to be 0.32, which is very close to the prediction in Eq.(\ref{beffscal}) for disks $3\sqrt{\pi}/16\simeq 0.33$}.
Note however that in this case, the asymptotic (algebraic) scaling regime is still not fully reached for the smallest solid fraction considered. As can be seen from Eq.(\ref{bn}), the speed of convergence is actually fixed by the parameter $x=p/k$, which accounts for the sparsity of the structure: convergence to the scaling regime is accordingly faster for more sparse structure with $x\gg 1$, {\it i.e.} small $d_f$, while $x=3/2$ in the present case.
%This scaling behavior can be checked to be qualitatively in 
%good agreement with the numerical results.
%Vleocity of convergence: x^n >> 1: x the smaller the faster the convergence to the scaling behavior. Here p/k

However, as we now show, a quantitative agreement can even be reached using the full expression in Eq.\ref{bn} to evaluate the slip length. 
%Again, one may reach a quantitative agreement by fixing the prefactor for this prediction to a calculated value for the slip length at low iteration.
We use  the numerical value for $b_2$ at the $n=2$ iteration (with $\phi_{s2}=0.05$) as an input parameter to fix the prefactor for this law in order to compare to numerical results. With this reference value, the slip length takes the following expression
$b_{\rm eff}(\phi_s)=b_2\times (x^n-1)/(x^2-1)$, with $n=\log(\phi_s)/\log(\phi_0)$.
%Thi corresponding to a surface fraction $\phi_{s2}=0.05$, and. 
As shown in Fig.\ref{Fig3}, this prediction is in quantitative agreement with the numerical results. 
%As for the case $x=1$ above, 
Note that if one uses the $n=1$ iteration as a reference for the prefactor parameter, the agreement is slightly less good, %due to the fact that $n=1$ corresponds to a rather high value of $\phi_s$, 
as the scaling approach gives only a fair, semi-quantitative estimate to the slip length $b_1$.

 % due to the neglect of hydrodynamic interactions. 
%is not taken into account in the hierarchical approach, the model is less good to describe the divergence
%Again this is due to the fact that the incremental law does not take into account the hydrodynamics interaction.\\

%We plotted in Fig.\ref{Fig3} the asymptotic scaling resulg 
%{\color{red}ÊFinally, we note that in this regime with low fractal dimension, the prefactor of the scaling law can be estimated to a good accuracy from
%Eq.(\ref{beffbis}) (assuming that the prefactor for circular disks is close to the one for square posts). FIGURE ?}

\par

Altogether the numerical results do confirm the theoretical predictions. A quantitative agreement can even be obtained by fixing the prefactor of the scaling behavior, using the numerical value of the slip length of the elementary structure (at the iteration $n=1$ or $n=2$).
\vskip0.5cm

\begin{figure}
    \centering
        \includegraphics[width=0.45\textwidth]{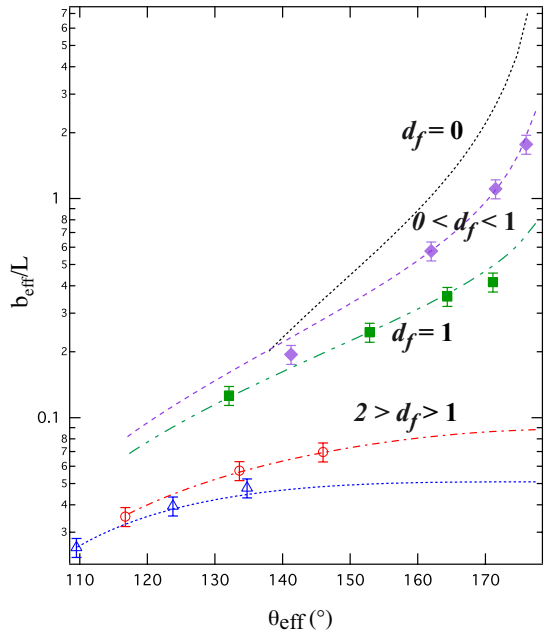}
    \caption{Normalized effective slip length $b_{\rm eff}/L$ in logarithmic scales, as a function of the effective contact angle $\theta_{\rm eff}$ for different  2D SH surfaces.  Symbols are the same as in Fig.\ref{Fig3}.
%    Numerical results for hierarchical fractal SH surfaces $\blacklozenge$:$k$=2,$p$=3, $\blacksquare$:$k$=3,$p$=3 ($d_f<1$),  $\circ$:$k$=5,$p$=3 ($d_f=1$), $\triangle$: $k$=6,$p$=3 (2$>d_f>1$).  (...):  Incremental law for $k$=6,$p$=3  and (-.-): incremental law for $k$=5,$p$=3 
    The black dotted line (top) is the scaling law for SH surface made out of periodic posts as predicted from Ref.  \cite{Ybert2007}. }\label{Fig4}
\end{figure}

Finally we consider in this paragraph the relationship between slippage and contact angle.
The hierarchical  fractal SH surface is characterized by its effective contact angle as defined in Eq.\ref{thetaeff} for a given iteration number $n$. 
We have plotted in Fig.\ref{Fig4} the effective slip length as a function of $\theta_{\rm eff}$ for both hierarchical fractal SH surfaces and for SH surfaces made out of periodic posts (numerical results and scaling laws).  Here we defined the effective contact angle as $\theta_{\rm eff}=\textrm{acos}(\phi_s-1)$, thus assuming for illustration purposes that the contact angle on the corresponding smooth surface is equal to $\theta_0=90^{\circ}$.

This plot illustrates immediatly that the slip length is not a unique function of the contact angle $\theta_{\rm eff}$: in contrast, it depends strongly on the geometry of the underlying structure for a given value of $\theta_{\rm eff}$, as characterized here by various fractal dimensions $d_f$. The hydrodynamic properties of a SH surface and thus of a hierarchical fractal  SH surface can not be simply determined from its effective contact angle. A strongly non-wettable surface with a contact angle of $\theta_{\rm eff} \simeq 180^\circ$ may well exhibit a low slip length and thus be inefficient in terms of drag reduction.

\subsection{Numerical results for slippage over hierarchical 1D structures}

\begin{figure}
    \centering
        \includegraphics[width=0.45\textwidth]{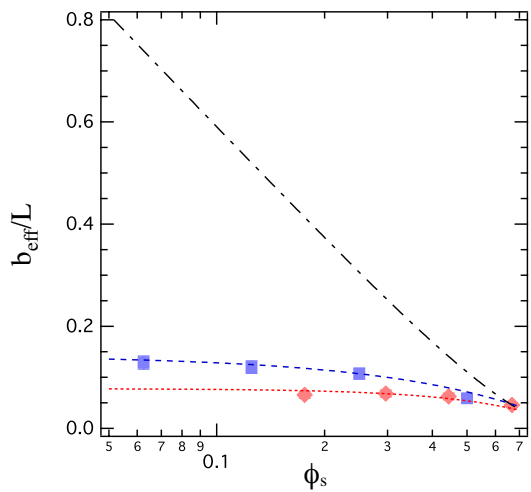}
    \caption{ {Comparison between the numerical results and the scaling law approach for 1D structures}: numerical results for the normalized slip length $b_{\rm eff}/L$ versus solid fraction $\phi_s$.
    %Comparison between the numerical results and the scaling law approach}. 
    %Normalized slip length $b/L$  as a function of the solid fraction $\phi_s$ 
    %for SH hierarchical fractal surfaces with 1D structures. 
    Theoretical predictions of the scaling approach, Eq. (\ref{bn}), are shown as lines for the two systems under consideration (see text for details).  Numerical results are from top to bottom:  \{$k$,$p$\}=\{2,4\} ($\blacksquare$);  \{$k$,$p$\}=\{2,3\} ($\blacklozenge$).The dash-dot  line is the analytical solution for a flow parallel to periodic grooves as predicted in Ref. \cite{PHILIP1972a}.\\
%    {\color{red} Pourquoi on ne peut pas estimer le reste des courbe theoriques a partir d'une valeur de reference pour la prediction ?} 
 }\label{Fig5}
\end{figure}

We finally turn to surfaces with 1D structures. 
%the comparison between the incremental law and the numerical results for . 
We have plotted on Fig.\ref{Fig5} the numerical results and the scaling laws for $b_{\rm eff}/L$ as a function of $\phi_s$ (in log scale) for 1D hierarchical fractal SH  surfaces characterized by different couples \{$k$,$p$\}. We have also plotted the analytical expression expected for a flow parallel to periodic grooves \cite{PHILIP1972a}. As for the 2D structures, the fractal surfaces yield values for the slip length which are far smaller than the one obtained on regular grooves. 
Furthermore as seen on this figure, we find a good agreement between the numerical results and the results of
the scaling law approach. As for the 2D structure above, the prefactors for the scaling relationships are obtained by matching the slip length for the $n=2$ or $n=3$ iteration, as the scaling approach is expected to be valid for relatively small values of $\phi_s$.
% if we consider the lower values of $\phi_s$ as in 2D. Indeed for high values of $\phi_s$ the hydrodynamic interactions are non negligible and we do not take them into consideration into our iterative model. Those interactions are taken into account in the numerical resolution. 
 
% We also underline the fact that the slip length quickly saturates for 1D hierarchical SH  fractal surfaces at low $\phi_s$. Those surfaces are not really efficient as regards to slip length and it is not worth fabricating 1D fractal SH surfaces with many iterations  for high slippage applications.\\

%\subsection{Discussion of the numerical results}

%\par

%We now comment on the results presented.
%

%In our 
%The only value that needs to be calculated numerically is $b_1$ the slip length after one iteration. Once this value is calculated the slip lengths for all iterations can be calculated using the scaling laws.

\par

%In 2D, for $d_f=1$ we recover with the scaling law approach a logarithmic divergence of the slip length ????une idee de justification?????.

%
%%%%%%%%%%%%%%%%%%%
% CONCLUSION
%%%%%%%%%%%%%%%%%%%
%

%\subsection{Stability versus slippage}

\section{conclusion}
In conclusion, we developed scaling laws for slippage on hierarchical fractal SH surfaces, providing explicit expressions for the slip length as a function of the solid fraction $\phi_s$ and fractal dimension $d_f$ of the surface.  For two dimensional structures, the slip length is found to scale as $b_{\rm eff} \sim \phi_s^{-\alpha}$ with $\alpha={{1-d_f}\over {2-d_f}}$ and $d_f$ the fractal dimension (here $d_f=\log k/\log p$ for the hierarchical surfaces under consideration).

We checked the validity of this approach by numerical calculation of the full hydrodynamic equations. 
The numerical results show that the simple scaling description remains valid up to relatively high values of solid fractions, even though the scaling laws do not take properly into account the hydrodynamic interactions. A quantitative agreement can be obtained by fixing the prefactor of the scaling behavior, using the numerical value of the slip length of the elementary structure (at an iteration $n=1$ or $n=2$).
For $d_f \le 1$ the scaling laws provide a good prediction for the slip length up to $\phi_s=0.1$ and for $d_f>1$ for which there is a fast saturation of the slip length, the agreement is good even for higher values of $\phi_s$. 
{Globally the numerical results confirm that the scaling approach captures the physical mechanisms at play: although hydrodynamic interaction are neglected between posts, scaling results are shown to be predictive over a
broad range of solid surface fraction.}

Altogether,
our study shows that for a given solid fraction $\phi_s$, fractal SH surfaces are less efficient than SH surfaces with regular pattern for drag reduction purposes. Furthermore
for a given effective contact angle $\theta_{\rm eff}$, the slip length on fractal SH surfaces  decreases  with increasing fractal dimension. %as compared to regular SH surfaces.  
%The value of $\theta_{\rm eff}$ is not enough to fully characterize the slippage on fractal surfaces. 
This shows that slippage depends strongly on the geometry of the underlying structure for a given value of $\theta_{\rm eff}$ and the amount of slippage on a fractal  SH surface can not just be determined from its effective contact angle. A strongly non-wettable surface with a contact angle close to  $180^\circ$ may well exhibit a low slip length and thus be totally inefficient in terms of drag reduction.
{This result is actually counter-intuitive: it shows that a surface with vanishingly small solid fraction leads to a finite friction force.}

%There is, in 2D,  a strong dependency of the slip length with the fractal dimension $d_f$. The scaling laws evaluate this slip length as a function of the fractal dimension of the surface.
%We also showed that  1D hierarchical fractal surfaces are not really efficient for slippage.
 The use of fractal SH surfaces appears however as a good compromise between the positive effect of stabilization of the Cassie state and a moderate reduction of the slippage effect. Very large slip length and corresponding large drag reduction can indeed be observed on regular SH surfaces, but usually at the expense of a low stability of the Cassie state versus pressure.  A recent study has demonstrated the recovery of the Cassie state after a transition towards the Wenzel state \cite{Lee2011}.  It would be thus interesting to consider surfaces with fractal structure in the context of stabilization issues \cite{Lee2009}. 
 
 Finally we mention that all results found in the present study for slip length can be immediately extended to heat transfer across fractal surfaces. In this case, the role of $b_{\rm eff}$ is played by the Kapitsa resistance \cite{Bocquet2007,Barrat2009}, which quantifies the temperature slip at the surface between two different materials. A key point indeed is that the scaling arguments proposed here rely ultimately on the Laplacian nature of Stokes equation, so that all scaling results presented here do apply equally to any transport property obeying a Laplacian equation, such as heat transport. Accordingly, one expects that the Kapitsa resistance (and corresponding Kapitsa length) at the interface with a hierarchical, fractal surface will obey similar scaling laws  versus the solid fraction as the ones found here for the slip length: the Kapitsa length is expected to scale as $\phi_s^{-\alpha}$ with $\alpha=(1-d_f)/(2-d_f)$ for surfaces with fractal dimension $d_f<1$ in the low $\phi_s$ regime, while it is expected to saturate at a finite value for $d_f >1$. This scaling results suggest therefore that for fractal dimensions $d_f>1$, it may be possible to have a relatively low Kapitsa surface resistance, thus maintaining a good heat transfer between the solid surface and the liquid, while keeping a strongly non-wetting superhydrophobic behavior. Such a compromise would be valuable for heat transfer problems in cooling pipes.
  
{As an extension of this study, our results could also be applied in a different field to the question of trapping of diffusing particles by patchy surfaces  \cite{Berez2004}, which is relevant to the questions of chemoreception and ligand binding
to surface receptors \cite{Berg1977,Szabo1982}. A counter-intuitive prediction emerging from our analysis
is that in the regime of an effective fractal dimension $1<d_{\rm eff} <2$, a hierarchical surface with {\it vanishingly small}  surface fraction of receptors may yet result in a {\it finite} effective trapping rate.}

% It would also be interesting in a future work to study those fractal surfaces as regards to their heat transfer potentialities, for which the equations are the same.

\begin{acknowledgments}
We thank Elisabeth Charlaix, Jean-Louis Barrat and Christophe Ybert for many discussions on the subject.
We acknowledge support from ANR, project {\it Mikado}. 
% and ERC Advanced Grant, project {\it Micromega}.
\end{acknowledgments}
\bibliographystyle{ieeetr}
%\bibliography{articlefrac}
%

\end{document}